\documentclass[sigconf, authorversion]{acmart}

\usepackage{booktabs} % For formal tables

\usepackage{hyperref}
\usepackage{url}
\newcommand{\quotations}[1]{``#1''}

\usepackage{enumitem}
\newlist{inparaenum}{enumerate}{2}% allow two levels of nesting in an enumerate-like environment
\setlist[inparaenum]{nosep}% compact spacing for all nesting levels
\setlist[inparaenum,1]{label=\bfseries\arabic*.}% labels for top level
\setlist[inparaenum,2]{label=\arabic{inparaenumi}\emph{\alph*})}% labels for second level

\usepackage{balance}
\usepackage{todonotes}

\begin{document}

\copyrightyear{2018}
\acmYear{2018}
\setcopyright{acmcopyright}
\acmConference[ARES 2018]{International Conference on Availability, Reliability and Security}{August 27--30, 2018}{Hamburg, Germany}
%\acmBooktitle{ARES 2018: International Conference on Availability, Reliability and Security, August 27--30, 2018, Hamburg, Germany}
\acmPrice{15.00}
\acmDOI{10.1145/3230833.3232807}
\acmISBN{978-1-4503-6448-5/18/08}

\title{Adding Salt to Pepper}
%\titlenote{}
\subtitle{A Structured Security Assessment over a Humanoid Robot}
%\subtitlenote{The full version of the author's guide is available as
  %\texttt{acmart.pdf} document}

\author{Alberto Giaretta}
\affiliation{%
  \department{Centre for Applied Autonomous Sensor Systems (AASS)}
  \institution{\"Orebro University}
  \country{Sweden}
}
\email{alberto.giaretta@oru.se}

\author{Michele De Donno}
\affiliation{%
  \department{DTU Compute}
  \institution{Technical University of Denmark}
  \country{Denmark}
}
\email{mido@dtu.dk}

\author{Nicola Dragoni}
\affiliation{%
  \department{DTU Compute}
  \institution{Technical University of Denmark}
  \country{Denmark}
}
\affiliation{%
  \department{Centre for Applied Autonomous Sensor Systems (AASS)}
  \institution{\"Orebro University}
  \country{Sweden}
}
\email{ndra@dtu.dk}

\renewcommand{\shortauthors}{Alberto Giaretta et al.}

\begin{abstract}
The rise of connectivity, digitalization, robotics, and artificial intelligence (AI) is rapidly changing our society and shaping its future development.
% The rise of connectivity, digitalization, and robotics, combined with the race for bringing artificial intelligence into every aspect of our daily life, is rapidly changing our society and shaping its future development.
During this technological and societal revolution, security has been persistently neglected, yet a hacked robot can act as an insider threat in organizations, industries, public spaces, and private homes. In this paper, we perform a structured security assessment of Pepper, a commercial humanoid robot. Our analysis, composed by an automated and a manual part, points out a relevant number of security flaws that can be used to take over and command the robot.
%In this paper, we perform a detailed security assessment of Pepper, a commercial humanoid robot, and we show how it is extremely easy to take over and command it.
Furthermore, we suggest how these issues could be fixed, thus, avoided in the future. The very final aim of this work is to push the rise of the security level of IoT products before they are sold on the public market.
% Furthermore, we suggest how these issues could be easily fixed, thus, strongly avoided in the future, as we believe that such products should undergo an inflexible security evaluation, before even considering to sell them on the public market.
\end{abstract}



% We no longer use \terms command


\maketitle

% !TEX root = main.tex
\section{Introduction}\label{sec:intro}
In 1990 an Artificial Intelligence (AI) winter violently hit the field and took it to its lowest point in history. Since then, AI popularity and enthusiasm constantly grew, up to the point where AI start-ups were able to raise more than \$6 billions in 2017, according to Venture Scanner~\cite{AI_fundings}. Even though AI permeates many fields without being acknowledged to be AI (as Nick Bostrom advocated in 2006~\cite{Bostrom_2006}), it is most obviously associated with robotics. We are witnessing an explosion of domestic robots which, equipped with Internet of Things (IoT) connectivity, make our lives more convenient and joyful.

But all that glitters is not gold. While these devices capabilities and potentialities are stunning, there seems to be a shallow approach to the security requirements of devices which are meant to closely interact with human beings and populated environments. A hacked robot, used for instance in a private home or even worse in a public space, like an airport, can have tremendous consequences for the safety of human beings, especially when it is easy
%a breeze
to remotely turn it into a ``cyber and physical weapon'', exposing malicious behaviors.

It is a matter of fact that the Internet of Things security state has been under the microscope for long time~\cite{DraGiaMaz17}, and threats like Mirai~\cite{dedonno2017FEDCSIS, dedonno2017analysis} showed that Distributed Denial of Service (DDoS) attacks can impact critical infrastructures and, subsequently, threaten human beings. Nevertheless, Internet-enabled robots are extremely powerful IoT devices equipped with a substantial number of sensors, motile capabilities, and often robotic arms that can grab and handle things. All in all, these characteristics make robotic devices one of the most critical classes of IoT devices currently sold off the shelf.

%\paragraph{Contribution of the Paper}
In this paper, we perform a thorough security assessment over Pepper, a commercial human-shaped social robot by SoftBank Robotics. The assessment is conducted in two main phases: an automated assessment and a manual one.
The results of our experiments are critical. %alarming.
We were able to steal the user credentials, perform a privilege escalation, and steal data. Moreover, we found out that it is possible to physically command the robot without authentication, use it to spy people and, potentially, even directly harm them.

%If this sounds scary, the reader should be even more frightened by realizing that SoftBank overlooked well-established security best practices and countermeasures. This product is exposed to extremely basic, yet very dangerous, flaws which were easily preventable since the beginning.

The security flaws highlighted in this work are another evidence of the general trend of shallowness undertook by IoT manufacturers with regard to the security of their devices. Indeed, it often results in the overlook of a number of well-established security best practices and countermeasures that exposes the final product to basic, yet very dangerous, security flaws which could be easily preventable.
% As we highlight throughout the paper, our biggest concern is that the manufacturer proved an evident shallowness, with respect to the security assessment of their device.
We have the feeling that commercial robots get on the market too quickly, evolving from research frameworks to final products without enough prior security investigation.
% from the manufacturers.
Thus, our final goal is not only to report our findings, but also to highlight that such devices should undergo a %stricter
prudent security evaluation, before becoming commercial products.

Please note that the findings of this paper are the result of experiments conducted on two samples of Pepper, provided to two different universities for research purposes. Therefore, we cannot assert that all the results are applicable to every Pepper on the market.

%\todo[inline]{...}

\paragraph{Paper Outline}

The paper is organized as follows.
Section~\ref{sec:pepper} describes the testbed robot. Section~\ref{sec:related_work} discusses related work. Section~\ref{sec:automated_assessment} and Section~\ref{sec:manual_assessment} respectively present the automated and manual security assessments we performed on Pepper, and related outcomes. Section~\ref{sec:fix} proposes some mitigations to the discovered vulnerabilities, and how similar issues might be prevented in future products. Lastly, Section~\ref{sec:conclusion} wraps up our investigation and draws the conclusion.
% lesson that the reader should learn.

% !TEX root = main.tex
\section{What is Pepper?}
\label{sec:pepper}
Manufactured by SoftBank Robotics, formerly Aldebaran Robotics, Pepper is a human-shaped social robot, designed to engage people and interact with them, not to help with practical duties. Its main characteristic is the capability to infer basic human emotions and react accordingly: it will be joyful if the engaged human is happy, comforting if sad, and so on.

\begin{figure}[ht]
\centering
  \includegraphics[width=0.48\textwidth]{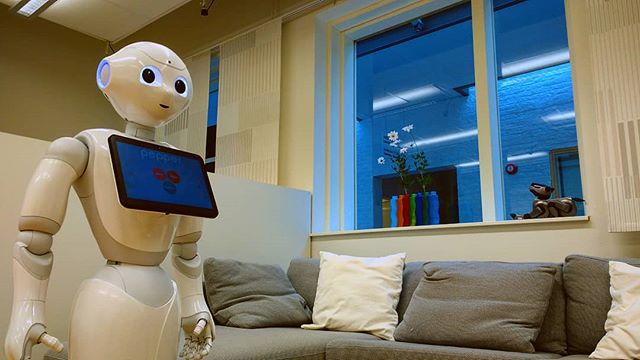}
  \caption{Pepper in one of the \"Orebro University robotic labs.}
\label{fig:pepper}
\end{figure}

Shown in Figure~\ref{fig:pepper}, Pepper strong points are perception and expression, and to achieve them the manufacturer equipped it with different sensors. Within its head lie 4 microphones, 2 HD cameras, a 3D depth sensor, and 3 touch sensors. Also, it has 2 more touch sensors, one per each hand, and 1 gyroscope in its torso. Last, the base is equipped with another gyroscope, 2 sonar sensors, 6 laser sensors, and 3 bumper sensors.

Currently, Pepper is used to welcome and inform people in various scenarios. In Japan, it is commonly seen in SoftBank Mobile stores and even in private houses. Allegedly, Nestl\'{e} is planning to equip more than 1000 Nescaf\'{e} sale points with a Pepper. Furthermore, early in 2017, a restaurant at Oakland Airport chose Pepper as an information tool and, around mid-February 2018, Lufthansa deployed it in Munich Terminal 2, as a first test phase aimed to evaluate whether passengers might accept it or not.

Moreover, Pepper can be operated through Robot Operating System (ROS)~\cite{ROS}, which is one of the most prevalent middlewares in robotics. Even though it is improperly called an Operating System (OS), ROS exposes many characteristics typical of OSes, such as hardware abstraction and low-level device control. Its collection of libraries and tools greatly simplifies the task of operating a large variety of robots that deeply diverge from each other from the hardware point of view.

% !TEX root = main.tex
\section{Related Work}
\label{sec:related_work}
To the best of our knowledge, there is no scientific paper in the literature discussing a structured security assessment of Pepper, nor of any other humanoid robot. However, for the sake of completeness, there are some works worth to mention.

First, the security consultancy company IOActive wrote a white paper~\cite{Cerrudo_2017} about a number of cyber-attacks performed over a set of different robots, amongst which Pepper. Their work gives an overview of the security vulnerabilities they found, and lists some generic recommendations that could be put in practice to improve the overall robotics security. Our study corroborates some of their results
%Their investigation confirms some of the results that we found
but, since they did not specifically focus on a single robot, their evaluation of Pepper is incomplete and many things are not analyzed nor reported. Indeed, our structured methodology helped us to depict a clearer overview about the general security state of this particular robot. %By following these clear steps, we discovered additional issues, such as lack of encryption that leaks passwords in cleartext, as well as a serious fragility against bruteforce attacks.

%the method used to achieve the goal is considerably different. On the one side, there is a white paper providing the list of security issues found on a pool of robots, using these vulnerabilities to suggest some best practices to improve the security of the robot industry. On the other side, there is a scientific manuscript taking the reader through the steps of a structured cyber-security assessment of a humanoid robot, ending up with possible countermeasures to its main vulnerabilities.\todo{rivedere questa estrema sintesi del nostro paper quando è completo}

Other works related to robotics security can be found in the literature. Clark et al.~\cite{Clark_2017} identify security threats in the robotics field, classify them based on the respective architectural layer, and discuss both their impact and possible countermeasures. Other works~\cite{Jeong_2017,White_2016} highlight and discuss a number of security issues of ROS (Robot Operating System). As already mentioned in Section~\ref{sec:pepper}, even though ROS is not necessary for Pepper to operate, it is a widespread middleware in the robotics community and compatible drivers are already available to command Pepper through ROS. This entails that ROS vulnerabilities can easily add up to the operated robot ones, and lead to an even worse situation.

%Although these manuscripts are loosely related to this paper, they have completely different focuses and results. In light of the considerations drawn so far, to the best of our knowledge, we present this work as the first structured cyber-security assessment of a humanoid robot.

% !TEX root = main.tex
\section{Automated Assessment}
\label{sec:automated_assessment}
The first step of our security assessment is to perform an automated analysis of the target.
% As a first step, we performed an automated security assessment over the robot.
This is a fundamental process to check the basic security level of any device and it is described in this Section.

First of all, we performed a port scanning. Secondly, we ran different tools to perform an automatic vulnerabilities scanning. All the information gathered in this process will be extremely useful in the subsequent manual phase.
The different steps and consequent results of the automated assessment are described in the rest of the Section.

\subsection{Port Scanning (Nmap)}
The aim of a port scanning is to discover the open network ports on a target, as well as the services bound to each open port. In our investigation, we used \textit{Nmap}~\cite{Nmap} to run a full port scanning, both for TCP and UDP ports. As shown in Table~\ref{tab:port_scan}, the manufacturer used out-of-date software, released around 2014.

\begin{table}[ht]
\caption{Nmap port scanning results.}
\begin{center}
\begin{tabular}{l@{\hskip 2em}l@{\hskip 2em}p{9em}@{\hskip 2em}l}
\hline
\rule{0pt}{12pt}Port & Protocol & Service & Released \\[2pt]
\hline\rule{0pt}{12pt}%
9 \hfill & mDNS & DNS-based service discovery & -\\
22 \hfill & SSH 2.0 & OpenSSH 6.6 & 2014\\
80  &     HTTP & nginx 1.4.7 & 2014\\
8002 & HTTP & Tornado httpd 3.1.1 & 2013\\
9559 & - & naoqi & -\\[2pt]
\hline
\end{tabular}
\end{center}
\label{tab:port_scan}
\end{table}

On the one hand, this is a benefit, because older software is usually more stable than the newest version. On the other hand, sticking to old software is a bad practice from a security point of view, since often the software team does not provide long term support, nor security updates, for outdated software. Therefore, in case a new vulnerability is discovered, old software has lower chances to receive updates, with respect to newer one. The most obvious example is Heartbleed~\cite{durumeric2014matter}, an implementation flaw found in dozens of versions of OpenSSL, around 2014. This bug exposed, for many years, private encryption keys of users to malicious attackers: a patch and new version of OpenSSL were immediately issued, and everyone was exhorted to immediately update the affected software.

Although a quick investigation on CVEdetails~\cite{CVEdetails} (the main information source for software vulnerabilities) shows that no critical security issues actually exist for the installed software packages, new vulnerabilities are discovered every day. As aforementioned, a system not regularly updated is potentially much more vulnerable to new exploits than a maintained one.

Apart from outdated software listening on standard port services, the port scanning showed an unusual naoqi service, binded to port 9559. We further investigate the matter in Section~\ref{subsec:remote_control}.

\subsection{Automated Vulnerabilities Scanning}
The aim of an automated vulnerabilities scanning is to discover if the robot services are vulnerable to well-known security issues.
We used two tools to perform the scanning: \textit{OpenVAS}~\cite{OpenVAS}, a complete vulnerabilities scanner framework, and \textit{OWASP ZAP}~\cite{OWASP_ZAP}, a web application vulnerabilities scanner. The results obtained by running these tools against Pepper are discussed below.

\subsubsection{OpenVAS}
The result obtained by executing OpenVAS against Pepper is highlighted in Table~\ref{tab:OpenVAS_result}. The tool raised 3 vulnerabilities and marked them accordingly to the severity: one as a medium and two as low.
\begin{table}[ht]
\caption{OpenVAS results. Severity scores are assigned by the tool itself.}
\begin{center}
\begin{tabular}{p{18em}@{\hskip 2em}l}
\hline
\rule{0pt}{12pt}Vulnerability & Severity \\[2pt]
\hline\rule{0pt}{12pt}%
SSH weak encryption algorithms supported & \textcolor{orange}{MEDIUM}\\
SSH weak MAC algorithms supported  &  \textcolor{cyan}{LOW}\\
TCP timestamp & \textcolor{cyan}{LOW}\\[2pt]
\hline
\end{tabular}
\end{center}
\label{tab:OpenVAS_result}
\end{table}
% \begin{itemize}
% 	\item SSH weak encryption algorithms supported, (MEDIUM severity);
% 	\item SSH weak MAC algorithms supported (LOW severity);
% 	\item TCP timestamp (LOW severity).
% \end{itemize}

The first warning is the troublesome one. Probably due to compatibility reasons, the manufacturer left enabled some known weak ciphers which could lead to third party exploitation. As an example, even though arcfour (RC4) is remarkably quick, many issues have been discovered throughout the years. Its usage led to severe problems in the past (e.g, RC4 is the main reason why WEP protocol turned out to be insecure) and it should not be used anymore, nowadays.

The second warning is related to the support of weak Message Authentication Code (MAC) algorithms.
% Some MAC algorithms, such as md5, are prone to collision attacks, where an attacker can create two different ssh keys that result in the same md5 hash. OpenVAS marks this as a low issue, since collision attacks are impractical for SSH keys, yet it is highly recommended to switch to SHA-256.
Today, some hash algorithms, such as MD5 and the ones producing a 96-bit hash value, are considered weak and thus not suitable to generate MACs \cite{CERT_2008}. Indeed, due to their proneness to hash functions collision attacks, it is possible to exploit them to conduct SSH downgrade attacks \cite{Bhargavan_2016}. Therefore, it is recommended to deprecate such algorithms and rely upon collision-resistant algorithms, such as SHA-2.

The third, and last, warning is a minor issue. Potentially, a malicious third party could estimate the target uptime and infer, with time, if the system is regularly patched (which often entails a reboot). Moreover, an attacker could evaluate whether a Denial of Service (DoS) attack was successful or not, again, by gathering the target uptime.

\subsubsection{OWASP ZAP}
\label{subsec:owasp}
The result obtained by executing OWASP ZAP against Pepper is highlighted in Table~\ref{tab:OWASP_ZAP_result}. The tool raised 3 warnings and marked them accordingly to the severity: one as a medium and the other two as low.
\begin{table}[ht]
\caption{OWASP ZAP results. Severity scores are assigned by the tool itself.}
\begin{center}
\begin{tabular}{p{18em}@{\hskip 2em}l}
\hline
\rule{0pt}{12pt}Vulnerability & Severity\\[2pt]
\hline\rule{0pt}{12pt}%
X-Frame-Options Header Not Set & \textcolor{orange}{MEDIUM}\\
Web Browser XSS Protection Not Enabled  &  \textcolor{cyan}{LOW}\\
X-Content-Type-Options Header Missing & \textcolor{cyan}{LOW}\\[2pt]
\hline
\end{tabular}
\end{center}
\label{tab:OWASP_ZAP_result}
\end{table}
%
% \begin{itemize}
% \item X-Frame-Options Header Not Set;
% \item Web Browser XSS Protection Not Enabled;
% \item X-Content-Type-Options Header Missing.
% \end{itemize}

The first warning implies that users are potentially exposed to Clickjacking attacks, a category of malicious techniques that trick users into clicking objects disguised as innocuous, through strategical layering of opaque and transparent layers. As an example, a transparent layer can be overlapped to a login web page and used to intercept keystrokes of the users.

The second warning highlights that a simple, yet effective, directive to protect from XSS (Cross-site scripting) attacks is not enabled on the web page, which exposes clients to arbitrary execution of malicious code. This issue, together with the first one, can have huge impact over the general security status of a web service.

The third warning, again, indicates that users are potentially exposed to XSS attacks and malicious code execution.
The MIME-Sniffing is a feature that some browsers use to derive the correct data flow, in case that the contacted web server does not declare correctly the MIME type. Unfortunately, if the website accepts uploads from the users without thoroughly checking the data content, this feature can be exploited by a malicious user. First, the attacker disguises a \textit{.html} file as a \textit{.jpg} one by changing the extension, and uploads it to the server (\textit{.html} are usually not allowed, whereas \textit{.jpg} ones are). Then, the attacker tricks the victim into opening the \textit{.jpg} file. Now, th MIME-Sniffing feature of the victim browser discovers that the \textit{.jpg} file is, as a matter of fact, a \textit{.html} file, therefore the browser overrides the server MIME type and execute the file as a \textit{.html}. In Section~\ref{subsec:sam_app}, we will show that one of the Pepper default applications enables the users to upload arbitrary files.

% !TEX root = main.tex
\section{Manual Assessment}
\label{sec:manual_assessment}

In the previous Section we have shown how we performed the initial security assessment of Pepper and gathered some basic information. Based on such information, we started looking deeper into the listening services and the related communication flow. As tools, we used \textit{Ettercap}~\cite{Ettercap} to perform an ARP spoofing attack and then we analyzed the intercepted traffic through \textit{Wireshark}~\cite{Wireshark}. Furthermore, we used \textit{Hydra}~\cite{Hydra} to perform a dictionary bruteforce attack and \textit{SSLsplit}~\cite{SSLsplit} as a Man-in-the-Middle (MitM) tool.

For the sake of completeness, we also investigated the system through a \quotations{uname -a} terminal command, which resulted in the following information: \textit{\quotations{Linux Pepper 4.0.4-rt1-aldebaran-rt \#1 [..] i686 Intel(R) Atom(TM) CPU E3845 @ 1.91GHz}}. After a quick verification of the CPU model, we report that Pepper is prone both to Meltdown and Spectre attacks~\cite{kocher2018spectre}.

\subsection{ARP Spoofing and Traffic Analysis (Ettercap \& Wireshark)}
\begin{figure}[ht]
  \includegraphics[width=\linewidth]{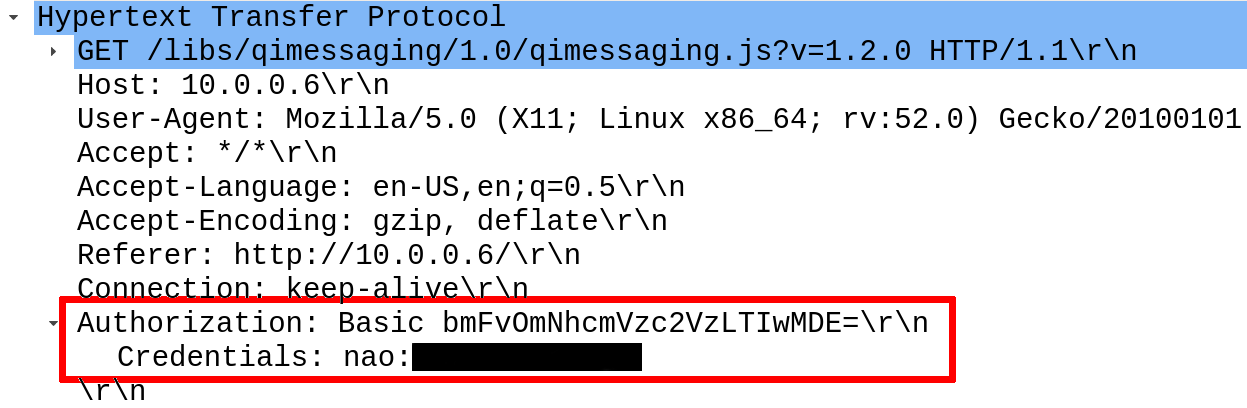}
  \caption{Authentication is performed over an unencrypted HTTP channel, easy to spoof for any attacker.}
\label{fig:http_sniffing}
\end{figure}

Pepper offers to users a nice web page where some administrative tasks, such as factory reset and password change, can be easily performed. As soon as we opened the web page, we noticed that the communication is established over HTTP, rather than HTTPS. Therefore, we decided to capture the login communication flow to steal the login credentials. As shown in Figure~\ref{fig:http_sniffing}, we were easily able to retrieve in clear text the \emph{user:password} pair, where \emph{user} is always \emph{nao}.

According to Pepper documentation~\cite{Pepper_Doc}, root login over SSH is disabled and only \textit{nao} user can perform a SSH connection, which is a positive thing from a security perspective. However, end-users can only set the password for \emph{nao} profile and no tool suggests, nor enables, them to change the root password. Besides, \textit{root} password is clearly written in the manual, and even easily guessable (i.e., \textit{root}). Unluckily, according to the documentation~\cite{Pepper_Doc}, the \emph{user:password} pair to log in the web application is the same for SSH. Therefore, after stealing the credentials, transmitted in clear text over HTTP, we logged as a normal user (i.e., \emph{nao}) via SSH and we issued a \emph{su} command to perform a privilege escalation.
In order to prevent this privilege escalation is sufficient to modify the root password. However, since there is no easy tool to change the password, the end-user should know how to issue a password change command through terminal, which is often not the case, thus, the chances that the root password will remain the default one are high.
% Since there is no easy tool to change root password, to prevent this privilege escalation, the end-user has to know how to issue a password change through terminal, and this complexity increases the chances that root password will remain the default one.

Moreover, we noticed that the web application does not provide any logoff feature, and this could be a potential problem. Indeed, in case that a user logs in from a public/shared workstation, either he makes sure to clear the browser cache, or he has no other way to logoff from the website.

As a remark, we want to stress that performing authentication over an unencrypted communication channel is a wrong and consistently dangerous security practice which should be always avoided, especially in commercial products.
% As a side note, we strongly believe that, in 2018, selling products so easily vulnerable to these kinds of attacks is not tolerable anymore. Authenticating over an unencrypted communication channel is one of the worst mistakes a software developer can make, computer scientists are taught to avoid it since their bachelor's program, it is quite saddening to find it implemented in a commercial product sold on the market.

\subsection{SSH Dictionary Bruteforce (Hydra)}
Knowing that the username \textit{nao} is fixed for this device, we investigated whether a bruteforce attack could be performed over SSH or not. In particular, we assessed if any kind of bruteforce protections have been deployed, such as IP addresses blacklist, upper bound of allowed parallel connections, and so on.

\begin{figure}[ht]
  \includegraphics[width=\linewidth]{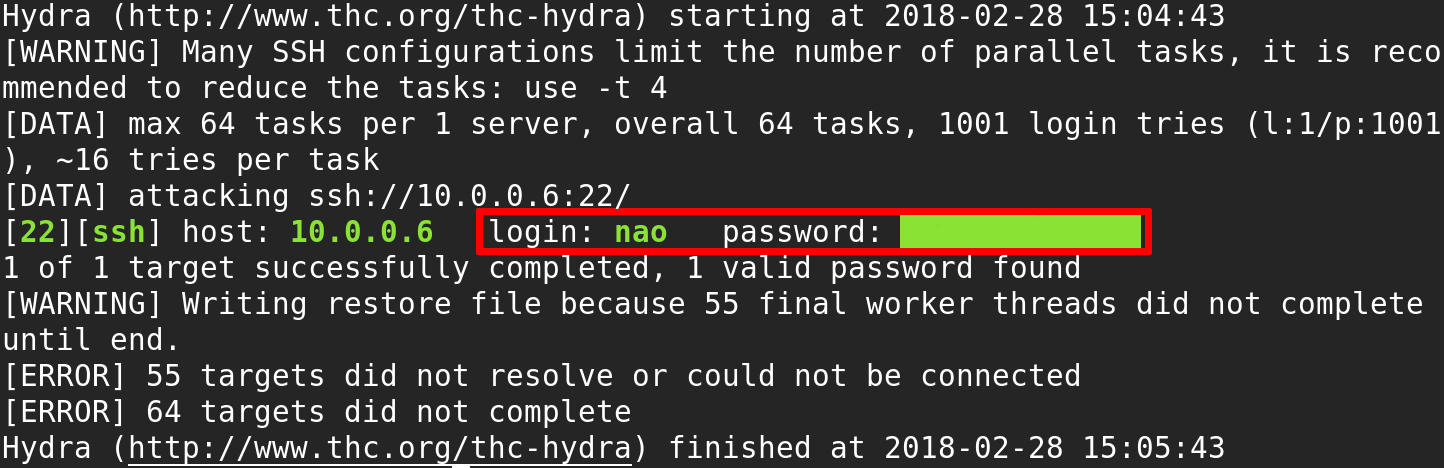}
  \caption{SSH dictionary bruteforce attack with Hydra. Comparing start and end time, Hydra was able to test 1000 entries in 1 minute (16 tries per second).%, achieving a respectable rate of more than 16 tries per second.
  }
\label{fig:hydra_ssh}
\end{figure}

In order to assess the presence of countermeasures, we used Hydra as a dictionary attack tool and we fed it with a 1000 entries dictionary. To ensure that bruteforce operations were carried on without glitches, as last entry of the dictionary we appended the real password, expecting to have a positive result at the end. We started by executing Hydra with 12 concurrent threads and the correct password was found in approximately 7 minutes, without experiencing any limitations. Then, we increased the concurrent threads to 16, and it took approximately 5 minutes to discover the password. Thus, we decided to double the threads and, again, we did not experience any limitations, confirmed by the fact that the password was found in approximately 2 minutes. Last, we used the maximum number of threads allowed by a single Hydra instance, which is 64 threads, and the password was found in exactly 1 minute, as shown in Figure~\ref{fig:hydra_ssh}.

This experiment clearly shows that no countermeasures to bruteforce attacks have been deployed with Pepper. %, which is once again an intolerable and disappointing finding.

For the sake of clarity, we chose to perform a dictionary attack to have a time-bounded result and assess whether bruteforce attacks are feasible or not. Nevertheless, the lack of countermeasures, and the consequent feasibility of the attack, holds even for pure permutation bruteforce attacks.

\subsection{Simple Animated Messages (SAM) Application}
\label{subsec:sam_app}
As previously discussed in Section~\ref{subsec:owasp}, the necessary condition to perform a MIME-Sniffing attack is the capability to upload a malicious script by changing its extension (e.g., from \textit{.sh} to \textit{.jpg}). Of all the default applications provided by Pepper, the Simple Animated Messages (SAM) application caught our attention. Once the user accesses the app web page at \path{<ip-address-pepper>/apps/sam/}, he can design a simple choreography that makes Pepper move, say something through a text-to-speech service, and show a picture on its on-board tablet. Moreover, the user can save the choreography and later edit it.

We were particularly interested in the picture upload feature, therefore we tried to upload a \textit{.jpg} file and we found out that the file is temporarily stored in the directory \path{/data/home/nao/.local/share/SAMService/uploads}. Once the user saves the choreography, the file is moved to \path{/data/home/nao/.local/share/SAMServ} \path{ice/backups/<Chor Name>}, where \path{<Chor Name>} is the name chosen by the user to save the choreography. Instead, if the user does not save the choreography, the directory \path{uploads} is emptied as soon as the application is closed.

Then, we tried to upload different files and we quickly figured out that the application performs no control over the file extension. As a matter of fact, we were able to upload images, text files which extensions have been modified to images, and even plain text files without performing extension editing. We did not perform a proof-of-concept attack, but it is evident %we have the strong feeling, to such an extent that we claim it,
that a malicious attacker can easily leverage this application to attack the machine of the administrator.

A possible attack scenario is the following. The attacker could create a script, save it as a \textit{.jpeg} file, and substitute the original file with the forged one. As a result, the on-board tablet of Pepper would stop to show the intended picture and, in order to fix this unexpected problem, the administrator could be led to access the SAM web page. At this point, given that no MIME-Sniffing protection is enabled, the browsers running on the administrator machine could interpret the forged file as a script and subsequently execute malicious arbitrary code.

\subsection{Man-in-the-Middle (SSLsplit)}
When it comes to secure communications, one of the critical points is how certificates are handled during the TLS/SSL handshake. As shown in~\cite{DBLP:conf/stm/ContiDG13}, wrong implementations of the TLS protocol can expose the communication to Man-in-the-Middle (MitM) attacks~\cite{ConDraLes2016}.

Analyzing Pepper communications through Wireshark, we noticed a number of encrypted packets, which led us to assess whether the handshake is carried on correctly or not. In order to do so, we used SSLsplit as a MitM tool and forged a self-signed certificate as similar as possible to the original one; after having ARP-spoofed both Pepper and the local gateway, we tried to feed the fake certificate to Pepper. These attempts were unsuccessful. Pepper seems to reject our fake certificates and keeps on refreshing the connection until it gets the correct certificate. This implies that the certificates validity is correctly assessed by Pepper and that a MitM attack is not easily done.

On a side note, ARP spoofing a target without performing a successful MitM attack usually cuts the target capability to connect to the Internet. However, after attacking Pepper, its standard applications kept on functioning even without Internet connection. Thus, to date, we cannot say exactly what kind of communication Pepper carries on the Internet, but we know that it does not seem to be vital for the basic operations of the robot.

\subsection{Remote Control without Authentication}
\label{subsec:remote_control}
As a strong point, Pepper exposes an API that enables users to program and command it. The APIs are available for a number of different programming languages, such as Python, C++, and Java. Users can operate the robot both through programs stored \& run on Pepper itself, and through execution of remote scripts. In the latter case, Pepper APIs enable users to access all the sensors, cameras, and microphones included, as well as, all the moving parts it has been equipped with, no exclusions.

While this functionality has interesting implications, we discovered that it is highly % astonishingly
insecure. Indeed, as previously said in Section~\ref{sec:automated_assessment}, Pepper exposes a service on port 9559 which accepts TCP messages and reacts accordingly. As long as the messages comply to the API, by-design, Pepper accepts packets from whoever sends them. In other words, we found out that anyone can write a simple script that remotely commands Pepper without providing any credential. Therefore, anyone can use Pepper to do a number of different things which include, but are not limited to, the following:
\begin{itemize}
\item Spy on people through Pepper cameras and microphones;
\item Interact with people and ask any kind of question, aiming to gather some personal information;
\item Hurt people by abruptly moving towards them or grabbing them;
\item Shut down and factory reset Pepper, through ALSystemProxy API module.
\end{itemize}
%Moreover, the ALSystemProxy API module enables anyone to shut Pepper down, factory reset it and so on.

In a research environment, the fact that no authentication is required to remotely control the robot could be a reasonable choice to keep things simple. However, as previously discussed in Section~\ref{sec:pepper}, Pepper is currently deployed in many scenarios where human beings are present and actively interact with it, and anyone can buy it off the shelf, thus, this security flaw can result in unpleasant outcomes. Moreover, it confirms once again the alarming trend that many robot manufacturers \cite{Cerrudo_2017} are undertaking by not considering the security as a key point in the design of their products.
% If this product were purely designed for research purposes, the very fact that no authentication is required it might be a less worrisome issue. However, as previously discussed in Section~\ref{sec:pepper}, Pepper is currently deployed in many scenarios where human beings are present and actively interact with it, and anyone can buy it off the shelf. This is why, again, we deem alarming the fact that, apparently, this manufacturer, along with many others~\cite{Cerrudo_2017}, invests almost nothing into constructing a secure product since the design phase.

% !TEX root = main.tex

\section{Results and Countermeasures}%Lesson Learned - How to Improve the Situation}
\label{sec:fix}
In this Section, we recap the steps we followed and the results obtained. In addition, we suggest some ideas to fix the issues we found and, consequently, to avoid the same problems in future products. The section follows the same order used in attacking the robot, from the initial port scanning to the remote control exploit.

\subsection{Wrapping Up: Steps and Outcomes}%Investigation Steps and Results}
As previously stated, we followed a structured methodology while conducting this security assessment, and we believe that others could benefit from the same kind of approach. Shown below, the steps that we deemed mandatory to have a clear and complete security overview of the analysed robot.
\begin{inparaenum}
  \item Automated assessments:
  \begin{inparaenum}
    \item Run a port scan software;
    \item Execute automated suites (e.g., OpenVAS).
  \end{inparaenum}
  \item Manual assessments:
  \begin{inparaenum}
    \item Analyse network traffic with Wireshark;
    \item Run bruteforce attacks;
    \item Assess SSL certificates handling;
    \item Manually investigate other interesting characteristics, such as uncommon open ports.
  \end{inparaenum}
\end{inparaenum}
\ \\
Following the aforementioned methodology we found a number of different vulnerabilities, briefly recapped in Table~\ref{tab:pepper_vulnerabilities}. Note that we manually assigned the severity marks in Table~\ref{tab:pepper_vulnerabilities}, according to our experience, whereas the scores in Table~\ref{tab:OpenVAS_result} and Table~\ref{tab:OWASP_ZAP_result} were automatically assigned from the respective tools.
\begin{table*}[ht]
\caption{Discovered Pepper vulnerabilities. We personally assigned the severity scores, according to our experience.}\label{tab:pepper_vulnerabilities}
\begin{center}
\begin{tabular}{l@{\hskip 2em}p{26em}@{\hskip 2em}l}
\hline
\rule{0pt}{12pt}Vulnerability & Practical Consequences & Severity\\[2pt]
\hline\rule{0pt}{12pt}%
Port scanning is not hindered & Easy to assess the attack surface & \textcolor{cyan}{LOW}\\
Software not updated & Higher exposition to exploits & \textcolor{orange}{MEDIUM}\\
Unrestricted file upload (in SAM) & Arbitrary code can be executed on the device & \textcolor{orange}{MEDIUM}\\
Admin web page over HTTP & Credentials in clear-text & \textcolor{red}{HIGH}\\
No bruteforce countermeasures & High exposure to password cracking & \textcolor{red}{HIGH}\\
Root password cannot be easily changed & High exposure to privilege escalation & \textcolor{red}{HIGH}\\
Unauthenticated remote control & Perform any action without authorization (e.g., take pictures) & \textcolor{red}{HIGH}\\[2pt]
\hline
\end{tabular}
\end{center}
\end{table*}

The practical implications of Table~\ref{tab:pepper_vulnerabilities} are critical for a device capable of doing as many things as Pepper. Stealing the admin credentials is really easy and, once in, getting root credentials takes just a command. From there, everything is possible and the only limit is the attacker skills; for instance an attacker can steal sensitive data or alter the communication flow. Even worse, due to design choices, a malicious attacker can remotely access to Pepper hardware without any kind of authorization. This entails that anyone could easily spy through Pepper cameras and microphones, as well as make it perform physical actions that could even harm human beings. Moreover, Pepper can potentially be used to hack other devices through its web services.

\subsection{Automated Assessment - Countermeasures}
The port scanning phase is fundamental to gather relevant information about the target: the more information, the more chances for the attacker to penetrate the system. One interesting way to counteract this technique is by using a software called Portspoof~\cite{Portspoof}, whose operation mode is counter-intuitive, yet really effective. Indeed, Portspoof opens all the 65535 TCP ports and returns a SYN+ACK reply to any incoming request. Moreover, for each TCP port that does not offer a service, Portspoof emulates a real service through fake replies and banners stored in a local database. As a result of these two combined techniques, port scanning becomes much less informative, since the attacker cannot distinguish between honest and fake services.

Next step, running some automated assessment tools is a good security practice %should be mandatory
for a new product. Once it has been done, the manufacturer can revise the vulnerabilities list and fix them. In our case, we found out that Pepper instance of OpenSSH supports weak ciphers, yet it takes a little effort to edit the configuration file \path{/etc/ssh/sshd\_config} and remove them. The same goes for the other warnings, which are clearly due to lack of personalization during the installation phases. In addition, what usually comes out from such assessments is that the software that run on similar devices are seriously outdated and this triggers a reflection. In our opinion, manufacturers are not only responsible to deliver a well-functioning product, but also to maintain it through regular security analyses and consequent updates.

\subsection{Manual Assessment - Countermeasures}
As already mentioned, %another example of the evident disregard of security implications,
the administration panel of Pepper runs over HTTP, leaking username and password in clear text. Even if it should be trivial, we consider important to remark, once again, that whenever sensitive information have to be communicated, HTTPS instead of HTTP is a must-have best practice. Communicating sensitive data over an insecure channel, such as HTTP, is strongly deprecated nowadays. %no longer tolerable nowadays.

Beside this, the fact that the \emph{root} password cannot be easily changed in Pepper deserves a comment of its own. On the one hand, it is true that remote root login is denied and an attacker has to know \emph{nao} password before escalate privileges. On the other hand, if the attacker finds out \emph{nao} password then root escalation cannot be prevented, since the user is never asked to change the root password. In recent past, we have seen that the same approach with credentials led to catastrophic results for the IoT world, with large-scale malware attacks such as Mirai~\cite{dedonno2017analysis, dedonno2017analysis}. Thus, we deem essential to enforce users to change all their devices passwords, the first time they power them on.

On a side note, nowadays one of the main problems with the username/pass\-word paradigm is that people have too many services and devices to administrate. This growing complexity leads to adopting weak password and reusing them on different platforms, which entails a low security level. To face this new challenge, we envision a future where the user/password paradigm is abandoned in favor of smarter Access Control (AC) mechanisms, based on blockchain smart contracts. As an example, UniquID is a blockhain access management startup, which aims to enable IoT devices to authenticate and communicate directly with each other, with no credentials needed, by means of a private blockchain~\cite{UniquID}.

Speaking about passwords, defending from remote bruteforce attacks is of paramount importance. There is a couple of easily applicable countermeasures to bruteforce attacks, as we already mentioned in Section~\ref{sec:manual_assessment}. First of all, IP blacklisting can help to cut out a single attacker that tries to infiltrate the device, but this is not sufficient. A motivated attacker, indeed, could use a botnet and coordinate a massive distributed bruteforce attack, in order to struck the attack from a large pool of IPs. Therefore, putting in place an upper bound of simultaneous connections can help, by strongly slowing down the tries per minute rate.

With respect to the SAM application, even though this is a non-critical application, it is useful to comment on this issue in order to address a more general problem. Whenever a user is allowed to upload a specific kind of file, whether it is a picture or any other type, the upload function should carefully analyse if the file adheres to the expected input. Controlling the file extension is not enough, and failing to filter wrong files can lead to arbitrary execution of code.

Another thing, manufacturers need to pay particular attention also to certificates handling, since MitM attacks can be highly disruptive. Even though our investigation showed that certificates are apparently handled correctly by Pepper, this is something that the manufacturer should  always verify and ensure for any new product.

Last, but definitely not least, a reflection about the robot remote operations is mandatory. We cannot define this a security bug, since an authentication process was not designed from the beginning, but rather an intrinsically insecure feature.  Nevertheless, this finding perfectly reflects the core message that we try to convey in this paper. Security assessments and features should be inherently part of a product life-cycle, not a layer that manufacturers impose on a finite product, in an attempt to make it somehow (in)secure.
%it was not even designed to be secure from the very beginning. But this is far from being a consolation. This incredibly overlooked flaw reflects perfectly the core message that we try to convey in this paper. Security assessments and features should be inherently part of a product life-cycle, not a layer that manufacturers impose on a finite product, in an attempt to make it somehow (in)secure.

To address this critical situation, the whole APIs should be redesigned in order to achieve good security standards, and it is clear to see that this would end up being not only expensive but also time consuming. Once again, the problem is that security is not a property that the average consumer values when buys commercial products, since users are generally not technically skilled enough to evaluate it. Therefore, unless it becomes a characteristic that people ask when they purchase an IoT device, the only way to enforce security is through regulations.

With Pepper, it seems that the manufacturer took a device designed to be used as a research framework and made a commercial product out of it. However, this is not the only case, Cerrudo et al.~\cite{Cerrudo_2017} showed that other products from different manufacturers have similar faults, and the authors express exactly our same concerns.
% since the beginning we had the feeling that manufacturer took a device designed to be used as a research framework and made a commercial product out of it. But this is not the only case, Cerrudo et al.~\cite{Cerrudo_2017} showed that other products from different manufacturers have similar faults, and the authors express exactly our same concerns.

% !TEX root = main.tex
\section{Conclusion}
\label{sec:conclusion}
In this paper, we performed a security assessment over a human-shaped social robot manufactured by SoftBank Robotics, namely Pepper. Our assessment, conducted through an automated phase and a manual one, pointed out a number of security flaws which are an evidence of the general trend of shallowness that many IoT manufacturers are undertaking with regard to the security status of their final products.
% In this paper, we performed a set of different security assessments, both automated and manual, over a human-shaped social robot manufactured by SoftBank Robotics, namely Pepper. In our investigation, we found a troublesome number of serious security flaws which exhibit that the manufacturer extensively neglected any sort of security assessments before commercializing their product.

The vulnerabilities we discovered enable an attacker to easily spoof login credentials, steal data stored in the robot, hack other connected devices that interact with it, and even physically harm human beings. In particular, we found out that Pepper APIs are inherently flawed and accept TCP packets from any unauthenticated source, assuming
%, with no apparent good reason, that no malicious attacker might call the same APIs.
that only legitimate users will call the APIs.

This case is a blatant example of our core message: manufacturers should consider security aspects of their products, before selling them on the market. Until now, traditional IoT devices were very simple, therefore their security flaws did not raise enough awareness about the consequent risks. Now, we are starting to deal with devices that cannot only jeopardize the security of human beings, but also their safety. Thus, overlooking security assessments during the system design phase can lead to very dangerous consequences, as well as expensive remedies.
%make extremely easy to take the robot over and command it, as well as use it to steal credentials and even hack other connected devices (such as Internet-of-Things devices).

\balance
\bibliographystyle{ACM-Reference-Format}
\bibliography{bibliography}

\end{document}